\documentclass[twocolumn,showpacs,preprintnumbers,amsmath,amssymb]{revtex4}
\usepackage{amsmath,amssymb,graphicx}  
\usepackage{bm}
\usepackage[percent]{overpic}
\usepackage{natbib}

\begin{document}


\title{Lorentz Dispersion Law from classical Hydrogen electron orbits \\ in AC electric field via geometric algebra}

\author{Uzziel Perez}
\affiliation{National Institute of Physics, University of the Philippines, \\ Diliman, Quezon City, Philippines}

\author{Angeleene S. Ang}
\affiliation{Ateneo de Manila University, Department of Physics, \\ Loyola Heights,Quezon City, Philippines 1108}
\email{angeleene.ang@gmail.com}

\author{Quirino M. Sugon, Jr.}
\affiliation{Manila Observatory, Upper Atmosphere Division, \\ Ateneo de Manila University Campus}
\altaffiliation[Also at ]{Ateneo de Manila University, Department of Physics, \\ Loyola Heights,Quezon City, Philippines 1108}

\author{Daniel J. McNamara}
\affiliation{Manila Observatory, Upper Atmosphere Division, \\ Ateneo de Manila University Campus}
\altaffiliation[Also at ]{Ateneo de Manila University, Department of Physics, \\ Loyola Heights,Quezon City, Philippines 1108}

\author{Akimasa Yoshikawa}
\affiliation{Department of Earth and Planetary Sciences, Faculty of Sciences, \\ Kyushu University, Fukuoka, Japan }


\date{\today}

\begin{abstract}
We studied the orbit of an electron revolving around an infinitely massive nucleus of a large classical Hydrogen atom subject to an AC electric field oscillating perpendicular to the electron's circular orbit. 
Using perturbation theory in geometric algebra, we show that the equation of motion of the electron perpendicular to the unperturbed orbital plane satisfies a forced simple harmonic oscillator equation found in Lorentz dispersion law in Optics. 
We show that even though we did not introduce a damping term, the initial orbital position and velocity of the electron results to a solution whose absorbed energies are finite at the dominant resonant frequency $\omega=\omega_0$; the electron slowly increases its amplitude of oscillation until it becomes ionized. 
We computed the average power absorbed by the electron both at the perturbing frequency and at the electron's orbital frequency. 
We graphed the trace of the angular momentum vector at different frequencies. 
We showed that at different perturbing frequencies, the angular momentum vector traces epicyclical patterns. 
\end{abstract}


\maketitle

\section{Introduction}


In standard optics texts, the position $x$ of an electron of charge $q$ and mass $m$ under the time-varying electric field $E = \mathcal{E}e^{i\omega t}$ of light is given by\cite{Jackson, zangwhale,kleitak}
\begin{equation}
\ddot{x}+\Gamma \dot{x} + \omega_0^2 x = \frac{q}{2m} E, 
\end{equation} 
where $\Gamma$ is the damping coefficient, $\omega_0$  is the natural frequency of oscillation of the electron. The complex solution $\tilde x$ to this equation is shown by Akhmanov and Nikitin \cite{Akhmanovp96} to be 
\begin{equation}
\tilde x = \frac{q}{m}\frac{1}{\omega^2_0 - \omega^2 + i\omega \Gamma}\mathcal{E},
\end{equation} 
so that the power absorbed by the atom is
\begin{equation}
\langle P \rangle = \langle qE\dot{x} \rangle = \frac{q^2}{2m}\frac{\omega^2\Gamma}{(\omega^2_0 - \omega^2)^2 +\omega^2\Gamma^2}|\mathcal{E}|^2, 
\end{equation}
which in complex space becomes 
\begin{equation}
P = \frac{e}{4}i\omega \left(\mathcal{E}^{*}\tilde x + \mathcal{E}\tilde x e^{2i\omega t}\right).
\end{equation} 
Notice that the damping term $\Gamma$ makes the power absorbed finite at the resonant frequency $\omega=\omega_0$. 

At present it is still not clear why an atom can be described as a simple harmonic oscillator subject to sinusoidal electric field of light, e.g. what is  responsible for the restoring force constant $k = m\omega^2_0$ and what is the cause of damping force $\Gamma\dot{x}$? 

In this paper we wish to show that a forced harmonic oscillator equation can arise for a large Hydrogen atom 
with a circular orbital of frequency $\omega_0$ subject to a linearly polarized light of frequency $\omega$ whose corresponding wavelength $2\pi c/\omega$ is much larger than the electron's orbital radius $r_0$, e.g. microwave frequencies, so that the phase of light is approximately the same at any point in the orbital path within a certain time period. That is, if the electron is initially in circular orbit in the $xy$-plane, the position $s$ of the electron along the $z$-axis is given by 
\begin{equation} \label{harmonicequation}
\ddot{s}+\bm{\omega_{o}^{2}}s=-\frac{qE}{m}\cos\left(\omega t\right),
\end{equation} 
as similarly given in Born and Wolf \cite{BornandWolfp91}. Even though this equation does not have a damping term, we shall show that the energy absorbed at the resonant frequency $\omega=\omega_0$ remains finite, provided we take into account the position $\mathbf{r}$ of the electron in 3D and use the vector form of the electrical energy dissipation expression \cite{Jackson}: 
\begin{equation}
P = q\mathbf{E}\cdot\mathbf{\dot{r}}.
\end{equation}
We shall show that $\langle P \rangle $ is finite at the resonant frequency $\omega = \omega_0$ even though there is no damping.

In 1974, Bayfield and Koch experimentally studied the ionization of hydrogen atoms under microwave frequencies \cite{Bayfield}. Since then, many tried to study the interaction of microwave radiation with classical hydrogen atom within the context of Rydberg atoms.\cite{hanken} 
Some authors studied the interaction with circularly polarized light \cite{Cole, Brunello, Farelly, JGriffiths, Gajda}, while others such as Leopold \cite{Leopold}, Grosfeld and Friedland \cite{Friedland} and Neishtadt \cite{Neishtadt} focused on the linearly polarized case.

For Leopold, his Hamiltonian is of the form:
\begin{equation}
H(\vec{r}, \vec{p}) = \frac{1}{2}p^2 -r^{-1} + zF_{max}\cos\omega t,
\end{equation}
and the equations are solved using Monte-Carlo techniques. For Grosfeld and Friedland, their Hamiltonian is of the form: 
\begin{equation}
H = \frac{1}{2}p^2 -r^{-1} + Z\mu\cos\Psi,
\end{equation}
where the frequency $\omega (t) = d\Psi/dt$ and the authors used action angle variables. This Hamiltonian is the same one used by Neishtadt and Vasiliev, except that the latter authors used Delaunay elements. 
 
In our work, we shall not use the Hamiltonian approach. Instead, we shall use the force equation 
\begin{equation}
\label{eq:ddot r is -qE_Coul - qE_pert expand}
\ddot{\mathbf r} = -\frac{kq^2}{m}\frac{\mathbf r}{|\mathbf r|^3} - \lambda\mathbf e_3\frac{q}{m}E_0\cos(\omega t + \varphi), 
\end{equation}
and use linear perturbation theory to simplify the equation to a simple harmonic oscillator equation in (\ref{harmonicequation}) for a motion perpendicular to the initial circular orbital plane of the electron. This method is simpler than those of the previous authors because the solution to the simple harmonic oscillator equation is well-known. Just as in the optical dispersion theory, we computed for the average power absorption by the atom and showed that it only depends on the $z$-coordinate as in the standard theory. \begin{equation}
\langle P\rangle_\tau = \frac{1}{\tau}\int_0^{\tau}P\,dt .
\end{equation}
We shall show that the orbit of the electrons at integral frequency ratios are similar to De Broglie waves, except that the oscillation is perpendicular to the electron's orbital plane.
 
We shall divide the paper into six sections. Section 1 is Introduction. In Section 2, we shall discuss the Geometric Algebra formalism applied to planar rotations. In Section 3, we shall describe the unperturbed circular orbit of the electron around the nucleus. After this, we shall introduce an oscillating electric field perturbation and derive the equations of motion of the electron's oscillation perpendicular to its orbital plane, using the geometric algebra framework in our previous paper on Copernican epicyclical orbits\cite{sugoncopernican}. 
In Section 4, we shall compute the electron's orbital angular momentum and determine its limiting form at the resonant frequency. In Section 5, we shall compute the electrical power dissipation of the electron and plot the results for different values of the ratio between the orbital and light frequencies. We shall show that the average power, either over the perturbing or orbital period, is approximately similar to the standard absorption resonance curve with finite peak. Finally, we graph the angular momentum of the electron at different frequency ratios and show that the angular momentum vector traces epicyclical patterns. 

\section{Geometric Algebra}


\subsection{Scalars, Vectors, Bivectors, and Trivectors}

In Clifford (Geometric) Algebra $\mathcal Cl_{3,0}$, also known as the Pauli Algebra, the product of the three unit vectors $\mathbf{e}_1$,$\mathbf{e}_2$, and $\mathbf{e}_3$ satisfies the orthonormality relation \cite{vold, hestenesoersted, geomaloptics}
\begin{equation} 
\label{eq:orthonormality}
\mathbf e_j \mathbf e_k + \mathbf e_k \mathbf e_j = 2 \delta_{jk},
\end{equation}
where $\delta_{jk}$ is the Kronecker delta function. 
In other words, the square of the length of the vectors is equal to one and the product of two perpendicular vectors anticommute.

Let $\mathbf a$ and $\mathbf b$ be two vectors spanned by $\mathbf e_1$, $\mathbf e_2$, and $\mathbf e_3$.  We can show that their product satisfies the Pauli identity\cite{baylis1,lounesto}
\begin{equation}
\mathbf a\mathbf b = \mathbf a\cdot\mathbf b + i(\mathbf a\times\mathbf b),
\end{equation}
where 
$i=\mathbf e_1\mathbf e_2\mathbf e_3$
is the unit trivector which behaves like an imaginary scalar that transforms vectors to bivectors. 
The Pauli identity states that the geometric product of two vectors is equal to the sum of their scalar dot product and their imaginary cross product.

%


\subsection{Exponential Function and Rotations}

Let $ i \mathbf e_3 \theta$ be the product of a bivector $ i \mathbf e_3 = \mathbf e_1 \mathbf e_2$ with the scalar $\theta$. 
Since the square of $ i \mathbf e_3 \theta$ is negative,
then the exponential of $i \mathbf e_3 \theta$ is given by Euler's theorem
\begin{equation}
\label{eq:e^itheta}
e^{ i \mathbf e_3  \theta  } = \cos\theta +  i \mathbf e_3 \sin\theta.
\end{equation} 
From this we can see that
\begin{subequations}
\begin{align}
\label{eq:cos theta exponential}
\cos\theta &=\frac{1}{2}(e^{ i \mathbf e_3  \theta  } + e^{ -i \mathbf e_3  \theta  }),\\
\label{eq:sin theta exponential}
\sin\theta &=\frac{1}{2 i \mathbf e_3 }(e^{ i \mathbf e_3  \theta  } - e^{ -i \mathbf e_3  \theta  }),
\end{align}
\end{subequations}
which are the known exponential definitions of cosine and sine functions.

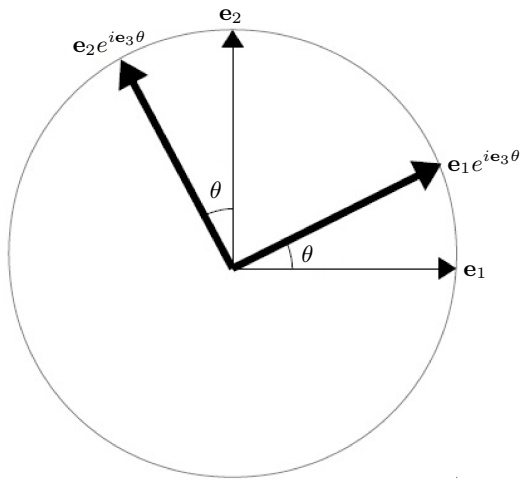
\begin{figure}[t!] 
\centering
\vspace*{1em}
   \begin{overpic}[
width=0.7\columnwidth,
tics=5,
page=1
]{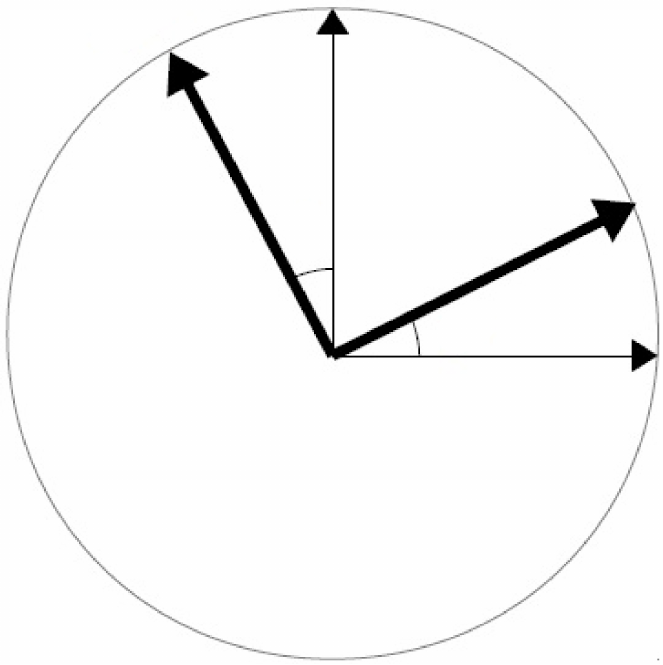}
     \put (97,68) {$\mathbf e_1 e^{i \mathbf e_3 \theta}$}
     \put (15,94) {$\mathbf e_2 e^{i \mathbf e_3 \theta}$}
     \put (47,101) {$\mathbf e_2$}
     \put (100.5,45) {$\mathbf e_1$}
     \put (65,48) {$\theta$}
     \put (45,62) {$\theta$}
  \end{overpic}
\caption{Rotation of $\mathbf{e}_1$ and $\mathbf{e}_2$ about $\mathbf{e}_3$ counterclockwise by an angle $\theta$}
\label{fig:garotation}
\end{figure}

Multiplying Eq.~(\ref{eq:e^itheta}) by $\mathbf e_1$, $\mathbf e_2$, and $\mathbf e_3$, we obtain
\begin{subequations}
\begin{align}
\label{eq:e_1e^itheta}
\mathbf{e}_1e^{ i \mathbf e_3  \theta }&= \mathbf{e}_{1}\cos\theta +\mathbf{e}_2\sin\theta = e^{- i \mathbf e_3  \theta }\mathbf{e}_1, \\
\label{eq:e_2e^itheta}
\mathbf{e}_2e^{ i \mathbf e_3  \theta }&= \mathbf{e}_{2}\cos\theta -\mathbf{e}_1\sin\theta = e^{- i \mathbf e_3  \theta }\mathbf{e}_2, \\
\mathbf{e}_3e^{ i \mathbf e_3  \theta }&= \mathbf{e}_{3}\cos\theta +\mathbf{e}_3  i \mathbf e_3 \sin\theta = e^{ i \mathbf e_3  \theta }\mathbf{e}_3 .
\end{align}
\end{subequations}
 Notice that $\mathbf{e}_1 e^{ i \mathbf e_3  \theta }$ is a rotation of $\mathbf{e}_1$ counterclockwise about $\mathbf{e}_3$ by an angle $\theta$, while  $\mathbf{e}_2 e^{ i \mathbf e_3  \theta }$ is a rotation of $\mathbf{e}_2$ counterclockwise about the same direction and the same angle. Notice, too, that the argument of the exponential changes sign when $\mathbf{e}_1$ or $\mathbf{e}_2$ trades places with the exponential, while $\mathbf{e}_3$ commutes with the exponential.

A vector $\mathbf a$ in 2D can be expressed in both rectangular and polar forms:
\begin{equation}
\label{eq:a is xe_1 + ye_2}
\mathbf a = a_x\mathbf e_1 + a_y\mathbf e_2  = a\mathbf e_1 e^{ i \mathbf e_3 \theta}.
\end{equation}
Expanding the exponential using Eq.~(\ref{eq:e_1e^itheta}) and separating the $\mathbf e_1$ and $\mathbf e_2$ components, we arrive at the standard transformation equations for polar to rectangular coordinates:
\begin{subequations}
\begin{align}
x &= a\cos\theta,\\
y &= a\sin\theta.
\end{align}
\end{subequations}

We may also factor out $\mathbf e_1$ in Eq.~(\ref{eq:a is xe_1 + ye_2}) either to the left or to the right to get
\begin{subequations}
\begin{align}
\mathbf a &= \mathbf e_1\hat a = \mathbf e_1(x +  i \mathbf e_3  y) = \mathbf e_1 ae^{ i \mathbf e_3 \theta},\\
\mathbf a &= \hat a^*\mathbf e_1 = (x -  i \mathbf e_3  y)\mathbf e_1 = ae^{- i \mathbf e_3 \theta}\mathbf e_1.
\end{align}
\end{subequations}
Factoring out $\mathbf e_1$ yields the definition of the complex number $\hat a$ and that of its complex conjugate $\hat a^*$:
\begin{subequations}
\begin{align}
\hat a &= a_x +  i \mathbf e_3  a_y = ae^{ i \mathbf e_3 \theta},\\
\hat a^* &= a_x -  i \mathbf e_3  a_y = ae^{- i \mathbf e_3 \theta}.
\end{align}
\end{subequations}
In general, we have the following relations:
\begin{subequations}
\begin{align}
\label{eq:e_1a is a*e_1}
\mathbf e_1\hat a &= \hat a^*\mathbf e_1,\\
\label{eq:e_2a is a*e_2}
\mathbf e_2\hat a &= \hat a^*\mathbf e_2,\\
\label{eq:e_3a is ae_3}
\mathbf e_3\hat a &= \hat a\mathbf e_3.
\end{align}
\end{subequations}
That is, $\mathbf e_1$ and $\mathbf e_2$ both changes the complex number $\hat a$ to its conjugate $\hat a^*$ after commutation, while $\mathbf e_3$ simply commutes with $\hat a$ \cite{jancewicz, geomaloptics, vold, hestenesoersted}.

\section{Light-Atom Interaction}

\subsection{Unperturbed Electron Orbit}

Classically, the position $\mathbf r$ of an electron of mass $m$ and charge $-q$ as it revolves around a massive proton of charge $q$ is given by Coulomb's law:
\begin{equation} \label{eq:Coulomb's law}
\ddot{\mathbf r} =-\frac{kq^{2}}{m}\frac{\mathbf r}{|\mathbf{r}|^{3}},
\end{equation} 
where $k$ is the electrostatic force constant. We claim that a solution to Eq. (\ref{eq:Coulomb's law}) is given by
\begin{equation}
\label{eq:r is e_1r_0psi_0} 
\mathbf{r}=\mathbf r_0 = \mathbf e_1\hat r_0\hat\psi_0,
\end{equation}
where 
\begin{subequations}
\begin{align}
\label{eq:r_0}
\hat r_0 &=r_0e^{ i \mathbf e_3 \varphi_0} \\
\label{eq:psi_0}
\hat\psi_0 &=e^{ i \mathbf e_3 \omega_0 t}
\end{align}
\end{subequations} 
are the complex amplitude and the rotation operator, respectively. Substituting these back to Eq.~(\ref{eq:r is e_1r_0psi_0}), we get
\begin{equation}
\mathbf r_0 = \mathbf e_1 r_0 e^{ i \mathbf e_3 (\omega_0 t +\varphi_0)},
\end{equation} 
which yields
\begin{align}
\label{eq:r is e_1r_0psi_0 rectangular}
\mathbf r_0 = \mathbf e_1 r_0\cos(\omega_0 t + \varphi_0) +\mathbf e_2 r_0\sin(\omega_0 t +\varphi_0),
\end{align}
after expanding the exponential and distributing $\mathbf e_1$.
Equation~(\ref{eq:r is e_1r_0psi_0}) states that the electron moving around the proton in circular orbit of radius $r_0$ with angular velocity $\omega_0$ and rotational phase angle $\varphi_0$.  

\begin{figure}
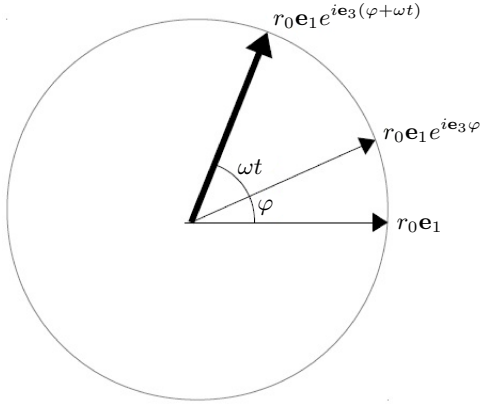

\centering
\vspace*{1em}
   \begin{overpic}[
width=0.6\columnwidth,
tics=5,
page=2
]{rydberg_version2_graphics}
     \put (69,97) {$r_0 \mathbf e_1 e^{i \mathbf e_3 (\varphi + \omega t)}$}
     \put (97,68) {$r_0 \mathbf e_1 e^{i \mathbf e_3 \varphi}$}
     \put (100.5,45) {$r_0 \mathbf e_1$}
     \put (65,50) {$\varphi$}
     \put (60,59) {$\omega t$}
  \end{overpic}
\caption{Uniform circular motion of an electron with a distance $r_0$ from the nucleus. The orbital angular frequency is $\omega$ and the phase angle is $\varphi$.}
\label{fig:gauniformcirc}
\end{figure}

To verify that Eq.~(\ref{eq:r is e_1r_0psi_0}) is indeed a solution to the Coulomb's law in Eq.~(\ref{eq:Coulomb's law}), we first compute the first and second time derivatives of Eq.~(\ref{eq:r is e_1r_0psi_0}):
\begin{subequations}
\begin{align}
\label{eq:dot r is e_1r_0iw_0psi_0}
\dot{\mathbf{r}} &=\mathbf e_1 \hat{r}_0 i \mathbf e_3 \omega_{o}\hat{\psi}_0,\\
\label{eq:ddot r is -e_1r_0w_0^2psi_0}
\ddot{\mathbf{r}} &=-\mathbf e_1\hat{r}_0\omega_{0}^{2}\hat{\psi}_0.
\end{align}
\end{subequations}
Now, substituting Eqs.~(\ref{eq:r is e_1r_0psi_0}) and (\ref{eq:ddot r is -e_1r_0w_0^2psi_0}) to the Coulomb's law in Eq.~(\ref{eq:Coulomb's law}), we obtain 
\begin{equation}
\label{eq:omega_0^2}
\omega_{o}^{2}=\frac{kq^2}{mr_0^3},
\end{equation}
after cancelling out $\mathbf e_1\hat r_0\hat\psi_0$.  Equation~(\ref{eq:omega_0^2}) is the familiar circular orbit condition.

\subsection{Perturbation by an Oscillating Field}

Suppose that the electron is subject not only to the Coulomb force $-q\mathbf E_{c}$ due to the proton, but also to the force $-q\mathbf E_{p}$ due to an oscillating perturbing field.  The equation of motion of the electron then becomes
\begin{equation}
\label{eq:ddot r is -qE_Coul - qE_pert}
m\ddot{\mathbf r} = -q\mathbf E_{c} - \lambda q\mathbf E_{p},
\end{equation}
where $\lambda$ is a perturbation parameter that shall later be set equal to unity.  More specifically, we write
\begin{equation}
\label{eq:ddot r is -qE_Coul - qE_pert expand}
\ddot{\mathbf r} = -\frac{kq^2}{m}\frac{\mathbf r}{|\mathbf r|^3} - \lambda\mathbf e_3\frac{q}{m}E_0\cos(\omega t + \varphi),
\end{equation}
where $E_0$, $\omega$, and $\varphi$ are the amplitude, angular frequency, and phase of the perturbing electric field.  Our aim is to determine the position $\mathbf r$ of the electron that satisfies Eq.~(\ref{eq:ddot r is -qE_Coul - qE_pert expand}).

\begin{figure}
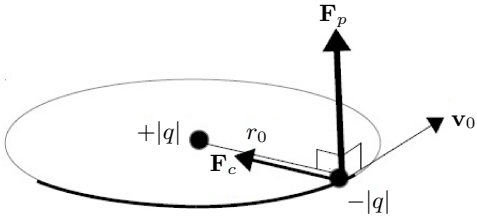

\centering
\vspace*{1em}
   \begin{overpic}[
width=0.7\columnwidth,
tics=5,
page=3
]{rydberg_version2_graphics}
     \put (31,17) {$+|q|$}
     \put (77,2) {$-|q|$}
     \put (47,10) {$\mathbf F_{c}$}     
     \put (71,43) {$\mathbf F_{p}$}     
     \put (100,20) {$\mathbf v_0$}     
     \put (55,17) {$r_0$}    
  \end{overpic}
\caption{The Coulomb force $\mathbf F_{c}$ and the perturbing force $\mathbf F_{p}$ on an electron moving with velocity $\mathbf v_0$ and radius $r_0$ }
\label{fig:1}
\end{figure}

To find the solution to the perturbed orbit equation in Eq.~(\ref{eq:ddot r is -qE_Coul - qE_pert expand}), we assume that the solution is a sum of the electron's unperturbed circular orbit in Eq.~(\ref{eq:r is e_1r_0psi_0}) and a slight perturbation $\hat s$ perpendicular to this orbit.  So we write
\begin{equation}
\label{eq:r is r_0 + lambda r_1}
\mathbf r = \mathbf r_0 + \lambda\mathbf r_1 = \mathbf e_1\hat r_0\hat\psi_0 +\lambda\mathbf e_3 \hat s.
\end{equation}
The first and second time derivatives of $\mathbf r$ are
\begin{subequations}
\begin{align}
\label{eq:dot r}
\dot{\mathbf r} &=\dot{\mathbf r}_0+\lambda\dot{\mathbf r}_1 = \mathbf e_1\hat r_0 i \mathbf e_3 \omega_0\hat\psi_0  +\lambda \mathbf{e}_3\dot{ \hat s},\\
\label{eq:ddot r}
\mathbf{\ddot r } &= \mathbf{\ddot r}_0 +\lambda\mathbf{\ddot r}_1 = -\mathbf e_1\hat r_0\omega_{o}^{2}\hat\psi_0+\lambda \mathbf e_3\ddot{ \hat s}.
\end{align}
\end{subequations}
Equation~(\ref{eq:ddot r}) shall take care of the left side of Eq.~(\ref{eq:ddot r is -qE_Coul - qE_pert expand}).

To expand the right-hand side, we need first to take the square of the position vector $\mathbf r$ in Eq.~(\ref{eq:r is r_0 + lambda r_1}) and retain only the terms up to first order in $\lambda$:
\begin{equation}
\label{eq:r^2 is r_0^2 + 2 lambda r_0 dot r_1}
\mathbf r^2 = \mathbf r_0^2+2\lambda(\mathbf r_0\cdot\mathbf r_1). 
\end{equation}
Since $\mathbf r_0$ lies on the unperturbed orbital plane of the electron $xy$ plane and $\mathbf r_1 = \mathbf e_3$ is perpendicular to this plane, then $\mathbf r_0\cdot\mathbf r_1=0$, so that Eq.~(\ref{eq:r^2 is r_0^2 + 2 lambda r_0 dot r_1}) reduces to
\begin{align}
\mathbf r^2 =\mathbf r_0^2 &= \mathbf e_1\hat r_0\hat\psi_0\,\mathbf e_1\hat r_0\hat\psi_0 = \mathbf e_1\mathbf e_1\hat r_0^*\hat\psi_0^*\hat r_0\hat\psi_0 \nonumber \\ 
&= \hat r_0^*\hat r_0 = r_0^2,
\end{align}
where we used the definitions of $\hat r_0$ and $\hat\psi_0$ in Eqs.~(\ref{eq:r_0}) and (\ref{eq:psi_0}). Thus, $|\mathbf r|=r_0$, so that
\begin{equation}
\label{eq:r/r^3}
\frac{\mathbf r}{|\mathbf r|^3} = \frac{1}{r_0^3}\mathbf e_1 \hat r_0\hat\psi_0.
\end{equation}
Equation~(\ref{eq:r/r^3}) shall take care of the Coulomb term on the right side of Eq.~(\ref{eq:ddot r is -qE_Coul - qE_pert expand}).

Now, substituting Eqs.~(\ref{eq:ddot r}) and (\ref{eq:r/r^3}) back to equation of motion in Eq.~(\ref{eq:ddot r is -qE_Coul - qE_pert expand}), we obtain
\begin{align}
-\mathbf e_1\hat r_0\omega_{o}^{2}\hat\psi_0+\lambda \mathbf e_3\ddot s &= 
-\omega_0^2(\mathbf e_1\hat r_0\hat\psi_0 + \lambda\mathbf e_3 \hat s) \nonumber \\&\quad \ - \lambda\mathbf e_3\frac{q}{m}E_0\cos(\omega t + \varphi),
\end{align}
where we used the circular orbit condition in Eq.~(\ref{eq:omega_0^2}). The term zeroth order in $\lambda$ cancels out, so we are left with the term first order in $\lambda$.  Hence,
\begin{equation}
\label{eq:ddot s + omega_0^2 s forced}
\ddot{ \hat s } + \omega_0^2 \hat s = -\frac{q}{m}E_0\cos(\omega t + \varphi),
\end{equation}
after rearranging the terms.  Notice that Eq.~(\ref{eq:ddot s + omega_0^2 s forced}) is a simple harmonic oscillator equation with sinusoidal forcing, which is the standard model for classical light-atom interaction.

\subsection{Solving the Forced Harmonic Oscillator Equation}

Let $\hat s_h$ and $\hat s_p$ be the homogeneous and particular solutions of the Eq.~(\ref{eq:ddot s + omega_0^2 s forced}).  That is,
\begin{equation}
\hat s = \hat s_h + \hat s_p,
\end{equation}
and
\begin{align}
\label{eq:s_h equation}
\ddot{\hat{s}}_h + \omega_0^2 \hat s_h &= 0,\\
\label{eq:s_p equation}
\ddot{\hat{s}}_p + \omega_0^2 \hat s_p &= -\frac{q}{m}E_0\cos(\omega t + \varphi).
\end{align}	 

The solution to the homogenous equation in Eq.~(\ref{eq:s_h equation}) is a sum of a sines and cosines:
\begin{equation}
\label{eq:s_h}
\hat s_h = c_{h1}\cos(\omega_0 t) + c_{h2}\sin(\omega_0 t),
\end{equation}
where $c_{h1}$ and $c_{h2}$ are scalar constants that will be determined from the boundary conditions.  On the other hand, the solution to the particular equation in Eq.~(\ref{eq:s_p equation}) is of the same form as the perturbing field:
\begin{equation}
\label{eq:s_p}
\hat s_p = c_p\cos(\omega t + \varphi),
\end{equation}
where $c_p$ is a scalar constant.  Substituting Eq.~(\ref{eq:s_p}) back to the particular equation in Eq.~(\ref{eq:s_p equation}) and solving for $c_p$, we get
\begin{equation}
c_p = \frac{qE_0}{m\omega_0^2}\frac{1}{(\alpha^2 - 1)},
\end{equation}
where
\begin{equation}
\alpha =\frac{\omega}{\omega_0}
\end{equation}
is the ratio of the perturbing frequency $\omega$ to the electron's orbital frequency $\omega_0$. Hence,
\begin{equation}
\hat s_p = \frac{q}{m\omega_0^2}\frac{1}{(\alpha^2 - 1)}E_0\cos(\alpha\omega_0 t + \varphi).
\end{equation}

Adding the homogenous solution $\hat s_h$ in Eq.~(\ref{eq:s_h}) to the particular solution $\hat s_p$ in Eq.~(\ref{eq:s_p}) yields the total solution:
\begin{align}
\label{eq:s expand}
\hat s &= c_{h1}\cos(\omega_0 t) + c_{h2}\sin(\omega_0 t) \nonumber \\
& \quad \ + \frac{q}{m\omega_0^2}\frac{1}{(\alpha^2 - 1)}E_0\cos(\alpha\omega_0 t + \varphi).
\end{align}
Its time derivative is
\begin{align}
\label{eq:dot s expand}
\dot{\hat s} &= -c_{h1}\omega_0\sin(\omega_0 t) + c_{h2}\omega_0\cos(\omega_0 t) \nonumber \\
&\quad \  - \frac{q}{m\omega_0}\frac{\alpha}{(\alpha^2 - 1)}E_0\sin(\alpha\omega_0 t + \varphi).
\end{align}

To determine the unknown constants $c_{h1}$ and $c_{h2}$, we first substitute the expressions for $s$ and $\dot s$ in Eqs.~(\ref{eq:s  expand}) and (\ref{eq:dot s expand}) back to the expressions for the position $\mathbf r$ and velocity $\dot{\mathbf r}$ in Eqs.~(\ref{eq:r is r_0 + lambda r_1}) and (\ref{eq:dot r}) to get
%

\begin{subequations}
\begin{align}
\label{eq:r with parameters}
\mathbf r &= \mathbf e_1\hat r_0\hat\psi_0 +\mathbf e_3(c_{h1}\cos(\omega_0 t) + c_{h2}\sin(\omega_0 t)) \nonumber\\
&\quad \  + \mathbf e_3 \frac{q}{m\omega_0^2}\frac{1}{(\alpha^2 - 1)}E_0\cos(\alpha\omega_0 t + \varphi),\\
\label{eq:dot r with parameters}
\dot{\mathbf r} &= \mathbf e_1\hat r_0 i \mathbf e_3 \omega_0\hat\psi_0 +\mathbf e_3(-c_{h1}\omega_0\sin(\omega_0 t) + c_{h2}\omega_0\cos(\omega_0 t)) \nonumber \\
&\quad \ - \mathbf e_3 \frac{q}{m\omega_0}\frac{\alpha}{(\alpha^2 - 1)}E_0\sin(\alpha\omega_0 t + \varphi),
\end{align}
\end{subequations}
after setting the perturbation parameter $\lambda = 1$.  If we assume that at $t=0$, the electron is in its unperturbed circular orbit around the nucleus, then
\begin{subequations}
\begin{align}
\mathbf r(0) &= \mathbf e_1\hat r_0,\\
\dot{\mathbf r}(0) &= \mathbf e_1\hat r_0 i \mathbf e_3 \omega_0.
\end{align}
\end{subequations}
Substituting these to Eqs.~(\ref{eq:r with parameters}) and (\ref{eq:dot r with parameters}), and setting $t=0$, we arrive at the expressions for the parameters $c_{h1}$ and $c_{h2}$:
\begin{subequations}
\begin{align}
\label{eq:c_h1}
c_{h1} &=  -\frac{q}{m\omega_0^2}\frac{1}{(\alpha^2 - 1)}E_0\cos\varphi,\\
\label{eq:c_h2}
c_{h2} &= \frac{q}{m\omega_0^2}\frac{1}{(\alpha^2 - 1)}E_0\sin\varphi .
\end{align}
\end{subequations}

Substituting Eqs.~(\ref{eq:c_h1}) and (\ref{eq:c_h2}) back to the expression for the position $\mathbf r$ in Eq.~(\ref{eq:r with parameters}), we get
\begin{align}
\label{eq:r parameters substitute}
\mathbf r &= \mathbf e_1\hat r_0\hat\psi_0 + \mathbf e_3\frac{qE_0}{m\omega_0^2}\frac{1}{(\alpha^2 - 1)} (-\cos\varphi\cos(\omega_0t) \nonumber \\
&\quad \ +   \sin\varphi\sin(\omega_0t) + \cos(\alpha\omega_0 t +\varphi)).
\end{align}
Using the identity for the cosine of a sum of two angles, Eq.~(\ref{eq:r parameters substitute}) reduces to
\begin{align}
\label{eq:r final}
\mathbf r &= \mathbf e_1\hat r_0\hat\psi_0 + \mathbf e_3\frac{qE_0}{m\omega_0^2}\frac{1}{(\alpha^2 - 1)} \times \nonumber \\
&\quad \ (\cos(\alpha\omega_0 t +\varphi) - \cos(\omega_0t +\varphi)).
\end{align}
Its time derivative is 
\begin{align}
\label{eq:dot r final}
\dot{\mathbf r} &= \mathbf e_1\hat r_0 i \mathbf e_3 \omega_0\hat\psi_0 +\mathbf e_3\frac{qE_0}{m\omega_0}\frac{1}{(\alpha^2 - 1)} \times \nonumber \\
&\quad \ (-\alpha\sin(\alpha\omega_0 t + \varphi) + \sin(\omega_0 t + \varphi),
\end{align}
where we used the definition of $\alpha =\omega/\omega_0$.  Equations~(\ref{eq:r final}) and (\ref{eq:dot r final}) are the position and velocity of the electron initially orbiting at radius $r_0$, angular frequency $\omega_0$, and phase $\varphi_0$, and perturbed by an oscillating electric field with amplitude $E_0$, frequency $\omega$, and phase $\varphi$. 

\begin{figure*}
 \centering
  \includegraphics[page=4, scale=0.85]{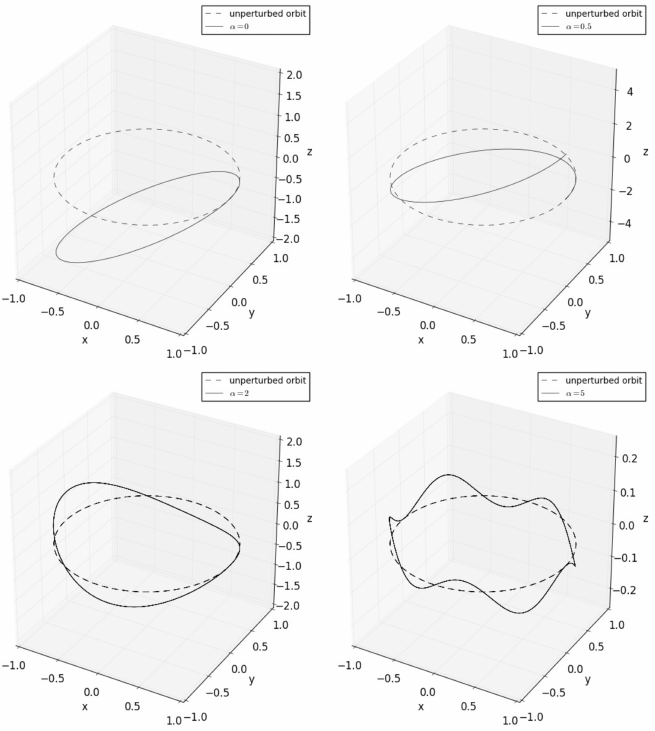}
\caption{The unperturbed orbit lies flat along the $\mathbf{e}_1-\mathbf{e}_2$ plane. When $\alpha = 0$, or when the electric field is constant in time, the orbit slants. When $\alpha \rightarrow \infty$, more waves are observed.}
\label{fig:large}
\end{figure*}

\subsection{Orbit Equations and Limiting Conditions}

To convert Eqs.~(\ref{eq:r final}) and (\ref{eq:dot r final}) into rectangular coordinates, we use the expansions in Eq.~(\ref{eq:e_1e^itheta}) and (\ref{eq:e_2e^itheta}), together with the identity $\mathbf e_1 i \mathbf e_3  = \mathbf e_2$ 
to arrive at
\begin{subequations}
\begin{align}
\label{eq:x final}
x &= r_0\cos(\omega_0 t + \varphi_0),\\
\label{eq:y final}
y &= r_0\sin(\omega_0 t + \varphi_0),\\
\label{eq:z final}
z &= \frac{qE_0}{m\omega_0^2}\frac{1}{(\alpha^2 - 1)}(\cos(\alpha\omega_0 t +\varphi) - \cos(\omega_0t +\varphi)), 
\end{align}
\end{subequations}
and 
\begin{subequations}
\begin{align}
\label{eq:dot x final}
\dot x &= -r_0\omega_0\sin(\omega_0 t + \varphi_0),\\
\label{eq:dot y final}
\dot y &= r_0\omega_0\cos(\omega_0 t + \varphi_0),\\
\label{eq:dot z final}
\dot z &= \frac{qE_0}{m\omega_0}\frac{1}{(\alpha^2 - 1)}(-\alpha\sin(\alpha\omega_0 t +\varphi) + \sin(\omega_0t +\varphi)).
\end{align}
\end{subequations}
Equations~(\ref{eq:x final}) to (\ref{eq:z final}) are the equations for plotting the orbit of the electron as a function of time.  Equations~(\ref{eq:dot x final}) to (\ref{eq:dot z final}) are for plotting the corresponding velocities.

When the perturbing frequency $\omega = 0$, corresponding to $\alpha=0$, the expressions for $z$ and $\dot z$ in Eqs.~(\ref{eq:z final}) and (\ref{eq:dot z final}) reduces to
\begin{subequations}
\begin{align}
z &= -\frac{qE_0}{m\omega_0^2}(\cos\varphi - \cos(\omega_0t +\varphi)),\\
\dot z &= -\frac{qE_0}{m\omega_0}\sin(\omega_0t +\varphi).
\end{align}
\end{subequations}
If $\varphi=0$, the perturbing field is $\mathbf E=E_0\mathbf e_3$, so that
\begin{subequations}
\begin{align}
z &= -\frac{qE_0}{m\omega_0^2}(1 - \cos(\omega_0t)),\\
\dot z &= -\frac{qE_0}{m\omega_0}\sin(\omega_0t).
\end{align}
\end{subequations}
On the other hand, if $\varphi=\pi$, the perturbing field is $\mathbf E=E_0\mathbf e_3$, so that
\begin{subequations}
\begin{align}
z &= \frac{qE_0}{m\omega_0^2}(1 + \cos(\omega_0t)),\\
\dot z &= \frac{qE_0}{m\omega_0}\sin(\omega_0t).
\end{align}
\end{subequations}
These are the behavior of the electron's orbit along the $z-$direction when the perturbing electric field is constant, also known as the DC electric field.

Now, when the perturbing frequency $\omega = \omega_0$, corresponding to $\alpha = 1$, the field resonates with the electron's orbit.  The only terms affected are $z$ and $\dot z$ in Eqs.~(\ref{eq:z final}) and (\ref{eq:dot z final}). Since both their numerators and denominators approach zero as $\alpha\rightarrow 1$, we apply L'hopital's rule by differentiating the numerators and denominators prior to evaluation of the limits:
\begin{subequations}
\begin{align}
\lim_{\alpha\rightarrow 1} z &= \frac{qE_0}{m\omega_0^2}\lim_{\alpha\rightarrow 1}\left(-\frac{\omega_0 t\sin(\alpha\omega_0 t)}{2\alpha}\right),\\
\lim_{\alpha\rightarrow 1} \dot z &= -\frac{qE_0}{m\omega_0}\lim_{\alpha\rightarrow 1}\left(\frac{\sin(\alpha\omega_0 t +\varphi)}{2\alpha} \right)\nonumber \\
& \quad \ -\frac{qE_0}{m\omega_0}\lim_{\alpha\rightarrow 1} \left(\frac{\alpha\omega_0 t\cos(\omega t + \varphi)}{2\alpha}\right).
\end{align}
\end{subequations}
Hence,
\begin{subequations}
\begin{align}
\lim_{\alpha\rightarrow 1} z &= -\frac{qE_0}{2m\omega_0}t\sin(\alpha\omega_0 t),\\
\lim_{\alpha\rightarrow 1} \dot z &= -\frac{qE_0}{2m\omega_0}(\sin(\omega_0 t + \varphi) +\omega_0 t\cos(\omega_0 t + \varphi)). 
\end{align}
\end{subequations}
Notice that the amplitude of the oscillations along $z$ and its corresponding velocity are linearly increasing in time.  Once the amplitudes of the oscillations becomes so large, our perturbation approximations breaks down.  Thus, our theory cannot really say what happens during ionization or whether ionization will really happen at all at the resonant frequency. (See Figs. \ref{fig:large} and \ref{fig:1}) 

\begin{figure}
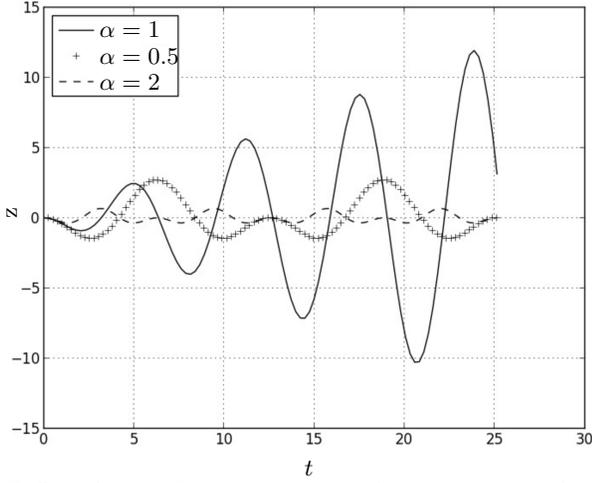

\centering
   \begin{overpic}[
width=0.9\columnwidth,
tics=5,
page=5
]{rydberg_version2_graphics}
     \put (15,70) {$\alpha=1$}
     \put (15,65.5) {$\alpha=0.5$}
     \put (15,61) {$\alpha=2$}
     \put (50,-5) {$t$}
     \put(0,40){\makebox(0,0){\rotatebox{90}{z}}}
  \end{overpic}
\caption{Height $z$ of the electron from its unperturbed circular orbit with respect to time.}
\label{fig:1}
\end{figure}

\section{Angular Momentum}

\subsection{Product Form}

Let us compute the product of the position $\mathbf r$ in Eq.~(\ref{eq:r is r_0 + lambda r_1}) and its velocity $\dot{\mathbf r}$ in Eq.~(\ref{eq:dot r}):
\begin{equation}
\mathbf r\dot{\mathbf r} = (\mathbf e_1\hat r_0\hat\psi_0 +\lambda\mathbf e_3 \hat s)(\mathbf e_1\hat r_0 i \mathbf e_3 \omega_0\hat\psi_0 +\lambda\mathbf e_3\dot{\hat s}).
\end{equation}
Distributing the terms, we get
\begin{align}
\label{eq:r dot r expand commute}
\mathbf r\dot{\mathbf r} &= \mathbf e_1\mathbf e_1\hat r_0^*\hat\psi_0^*\hat r_0 i \mathbf e_3 \omega_0\hat\psi_0  \nonumber \\
& \quad \ + \mathbf e_3\mathbf e_1 (\hat s\hat r_0 i \mathbf e_3 \omega_0\hat\psi_0 - \dot{\hat s}\hat r_0^*\hat\psi_0^*) + \mathbf e_3\mathbf e_3 s\dot{\hat s},
\end{align}
after setting the perturbation parameter $\lambda = 1$.  Since $i = \mathbf e_1\mathbf e_2\mathbf e_3$ and $ i \mathbf e_3  = \mathbf e_1\mathbf e_2$, then Eq.~(\ref{eq:r dot r expand commute}) reduces to
\begin{equation}
\label{eq:r dot r imaginary vector}
\mathbf r\dot{\mathbf r} = i\mathbf e_3\omega_0r_0^2 - i\mathbf e_1 s\hat r_0\omega_0\hat\psi_0 - i\mathbf e_2 s\hat r_0^*\hat\psi_0^* +\hat s\dot{\hat s} .
\end{equation}
Separating the scalar and bivector parts of Eq.~(\ref{eq:r dot r imaginary vector}), we arrive at
\begin{subequations}
\begin{align}
\mathbf r\cdot\dot{\mathbf r} &= \hat s \dot{\hat s},\\
\label{eq:r x dot r final}
\mathbf r\times\dot{\mathbf r} &= \mathbf e_3\omega_0r_0^2 - \mathbf e_1 s\hat r_0\omega_0\hat\psi_0 - \mathbf e_2 s\hat r_0^*\hat\psi_0^* ,
\end{align}
\end{subequations}
after factoring out the trivector $i$ in the second equation.  

Multiplying Eq.~(\ref{eq:r x dot r final}) by the electron's mass $m$ yields the the electron's angular momentum:
\begin{equation}
\label{eq:L psi}
\mathbf L = m\mathbf r\times\dot{\mathbf r} = \mathbf e_3\omega_0r_0^2 - \mathbf e_1 s \hat r_0\omega_0\hat\psi_0 - \mathbf e_2 s \hat r_0^*\hat\psi_0^*,
\end{equation}
where $s$ and $\dot{s}$ are the $\mathbf e_3$ components of the electron's position $\mathbf r$ and velocity $\dot{\mathbf r}$ in Eqs.~(\ref{eq:r final}) and (\ref{eq:dot r final}):
\begin{subequations}
\begin{align}
\label{eq:s final}
s &= \frac{qE_0}{m\omega_0^2}\frac{1}{(\alpha^2 - 1)}(\cos(\alpha\omega_0 t +\varphi) - \cos(\omega_0t +\varphi)),\\
\label{eq:dot s final}
\dot{s} &= \frac{qE_0}{m\omega_0}\frac{1}{(\alpha^2 - 1)}(-\alpha\sin(\alpha\omega_0 t + \varphi) + \sin(\omega_0 t + \varphi).
\end{align}
\end{subequations}
Using the definitions $\hat r_0 = r_0e^{ i \mathbf e_3 \varphi}$ and $\hat\psi_0 = e^{ i \mathbf e_3 \omega_0 t}$ in Eq.~(\ref{eq:L psi}), and separating the $\mathbf e_1$, $\mathbf e_2$, and $\mathbf e_3$ components, we arrive at
\begin{subequations}
\begin{align}
L_1 &= -m\omega_0 s r_0\cos(\omega_0 t + \varphi_0) - m\dot{s} r_0\sin(\omega_0 t + \varphi_0),\\
L_2 &= -m\omega_0 s r_0\sin(\omega_0 t + \varphi_0) - m\dot{s} r_0\cos(\omega_0 t + \varphi_0),\\
L_3 &= m\omega_0r_0^2,
\end{align}
\end{subequations}
which are the parametric expressions for the angular momentum in rectangular coordinates.

\subsection{Harmonic Form}

The vertical oscillation $s$ and its derivative $\dot{s}$ in Eqs.~(\ref{eq:s final}) and (\ref{eq:dot s final}) may be expressed in exponential forms:
\begin{subequations}
\begin{align}
\label{eq:s exponential}
s &= \frac{qE_0}{2m\omega_0^2}\frac{1}{(\alpha^2 - 1)}(e^{ i \mathbf e_3 (\alpha\omega_0 t +\varphi)} + e^{- i \mathbf e_3 (\alpha\omega_0 t +\varphi)}\nonumber \\ 
 & \qquad \qquad \qquad \quad - e^{ i \mathbf e_3 (\omega_0 t +\varphi)} - e^{- i \mathbf e_3 (\omega_0 t +\varphi)}),\\
\label{eq:dot s exponential}
\dot{s} &= \frac{qE_0}{2m\omega_0}\frac{- i \mathbf e_3 }{(\alpha^2 - 1)}(-\alpha e^{ i \mathbf e_3 (\alpha\omega_0 t +\varphi)} + \alpha e^{- i \mathbf e_3 (\alpha\omega_0 t +\varphi)} \nonumber \\
 &\qquad \qquad \qquad \quad \ + e^{ i \mathbf e_3 (\omega_0 t +\varphi)} - e^{- i \mathbf e_3 (\omega_0 t +\varphi)}) .
\end{align}
\end{subequations}
These may be rewritten as
\begin{subequations}
\begin{align}
\label{eq:s psi eta}
s &= \frac{qE_0}{2m\omega_0^2}\frac{1}{(\alpha^2 - 1)}(\hat\eta\hat\psi_0^\alpha +\hat\eta^*\hat\psi_0^{-\alpha}\nonumber \\ & \qquad \qquad \qquad \qquad -\hat\eta\hat\psi_0 -\hat\eta^*\hat\psi_0^{-1}),\\
\label{eq:dot s psi eta}
\dot s &= \frac{qE_0}{2m\omega_0}\frac{- i \mathbf e_3 }{(\alpha^2 - 1)}(-\alpha\hat\eta\hat\psi_0^\alpha + \alpha\hat\eta^*\hat\psi_0^{-\alpha} \nonumber \\
& \qquad \qquad \qquad \qquad + \hat\eta\hat\psi_0 - \hat\eta^*\hat\psi_0^{-1}),
\end{align}
\end{subequations}
where 
\begin{equation}
\hat\eta = e^{ i \mathbf e_3 \varphi}.
\end{equation}

Substituting Eqs.~(\ref{eq:s psi eta}) and (\ref{eq:dot s psi eta}) back to Eq.~(\ref{eq:L psi}) and noting that $\mathbf e_2 i \mathbf e_3  = -\mathbf e_1$, we obtain
\begin{equation}
\label{eq:L psi_L}
\mathbf L =  \mathbf e_3 m\omega_0r_0^2 - \mathbf e_1\frac{qE_0}{2\omega_0}\frac{1}{(\alpha^2 - 1)}\hat\psi_L,
\end{equation}
where 
\begin{align}
\label{eq:psi_L}
\hat\psi_L &= \hat\eta\hat r_0\hat\psi_0^{\alpha+1} + \hat\eta^*\hat r_0\hat\psi_0^{-\alpha+1} - \hat\eta\hat r_0\hat\psi_0^2 - \hat\eta^*\hat r_0\nonumber\\
&\quad\ -\alpha\hat\eta\hat r_0^*\hat\psi_0^{\alpha-1} + \alpha\hat\eta^*\hat r_0^*\hat\psi_0^{-\alpha-1} + \hat\eta\hat r_0^* - \hat\eta^*\hat r_0^*\hat\psi_0^{-2}.
\end{align}
Expanding the terms of $\hat\psi_L$ into exponential form, we get 
\begin{align}
\hat\psi_L &= r_0(e^{ i \mathbf e_3 ((\alpha + 1)\omega_0 t +\varphi +\varphi_0)} + e^{ i \mathbf e_3 ((-\alpha + 1)\omega_0 t -\varphi +\varphi_0)} \nonumber \\ 
& \qquad - e^{ i \mathbf e_3 (2\omega_0 t +\varphi +\varphi_0)}  - e^{ i \mathbf e_3 (-\varphi +\varphi_0)}\nonumber\\
&\qquad\ -\, \alpha e^{ i \mathbf e_3 ((\alpha - 1)\omega_0 t +\varphi -\varphi_0)} + \alpha e^{ i \mathbf e_3 (-(\alpha + 1)\omega_0 t -\varphi -\varphi_0)} \nonumber \\
&\qquad + e^{ i \mathbf e_3 (\varphi -\varphi_0)} - e^{ i \mathbf e_3 (-2\omega_0 t -\varphi -\varphi_0)}),
\end{align}
Substituting the result back to Eq.~(\ref{eq:L psi_L}) and separating the $\mathbf e_1$, $\mathbf e_2$, and $\mathbf e_3$ components, we arrive at
\begin{subequations}
\begin{align}
L_1 &= -\frac{qE_0r_0}{2\omega_0}\frac{1}{(\alpha^2 - 1)}\gamma_1,\\
L_2 &= -\frac{qE_0r_0}{2\omega_0}\frac{1}{(\alpha^2 - 1)}\gamma_2,\\
L_3 &= m\omega_0r_0^2,
\end{align}
\end{subequations}
where
\begin{subequations}
\begin{align}
\gamma_1 &= (1 +\alpha)\cos((\alpha + 1)\omega_0 t +\varphi +\varphi_0) \nonumber \\
&\quad+ (1 - \alpha)\cos((\alpha - 1)\omega_0 t +\varphi -\varphi_0)\nonumber\\
&\quad\ - 2\cos(2\omega_0 t +\varphi +\varphi_0),\\
\gamma_2 &= (1 -\alpha)\sin((\alpha + 1)\omega_0 t +\varphi +\varphi_0) \nonumber \\
&\quad - (1 + \alpha)\sin((\alpha - 1)\omega_0 t +\varphi -\varphi_0)\nonumber\\
&\quad\ + 2\sin(\varphi - \varphi_0).
\end{align}
\end{subequations}
Thus, since $\alpha = \omega/\omega_0$, we see that the orbit of the tip of the angular momentum vector $\mathbf L$ is a linear combination of circular motions with the following orbital frequencies: 
\begin{equation}
\omega_L = \{\pm(\omega +\omega_0), \pm(\omega - \omega_0), \pm 2\omega_0, 0\}.
\end{equation}

\subsection{Limiting Conditions}

In the DC field limit, $\alpha\rightarrow 0$, so that 
\begin{subequations}
\begin{align}
\lim_{\alpha\rightarrow 0} L_1 &= \frac{qE_0r_0}{2\omega_0}(\cos(\omega_0 t +\varphi +\varphi_0) \nonumber \\
&\qquad \qquad+ \cos(-\omega_0 t +\varphi -\varphi_0) \nonumber \\
&\qquad \qquad - 2\cos(2\omega_0 t +\varphi +\varphi_0)),\\
\lim_{\alpha\rightarrow 0} L_2 &= \frac{qE_0r_0}{2\omega_0}(\sin(\omega_0 t +\varphi +\varphi_0) \nonumber\\
&\qquad \qquad- \sin(-\omega_0 t +\varphi -\varphi_0) \nonumber \\
&\qquad \qquad+ 2\sin(\varphi - \varphi_0)).
\end{align}
\end{subequations}
On the other hand, in the resonance frequency limit, $\alpha\rightarrow 1$, both the numerators and denominators of $L_1$ and $L_2$ approach zero, so that we apply the L'hopital's rule:
\begin{subequations}
\begin{align}
\lim_{\alpha\rightarrow 0} L_1 &= -\frac{qE_0r_0}{4\omega_0}(\cos(2\omega_0 t +\varphi +\varphi_0) \nonumber \\
&\qquad- 2\omega_0 t\sin(2\omega_0 t +\varphi +\varphi_0) - \cos(\varphi -\varphi_0)),\\
\lim_{\alpha\rightarrow 0} L_2 &= -\frac{qE_0r_0}{4\omega_0}(-\sin(2\omega_0 t +\varphi +\varphi_0) \nonumber \\
&\qquad- \sin(\varphi -\varphi_0) - 2\omega_0 t\cos(\varphi - \varphi_0)) .
\end{align}
\end{subequations}
Notice that at the resonant frequency $\omega = \omega_0$, the $L_1$ and $L_2$ components of the angular momentum increases in time; in our perturbative approximation, the atom will be ionized.

\section{Power and Energy Absorption}

\subsection{Work-Energy Theorem}

\begin{figure*}
 \centering
 \begin{tabular}{cc}
  \includegraphics[page=6]{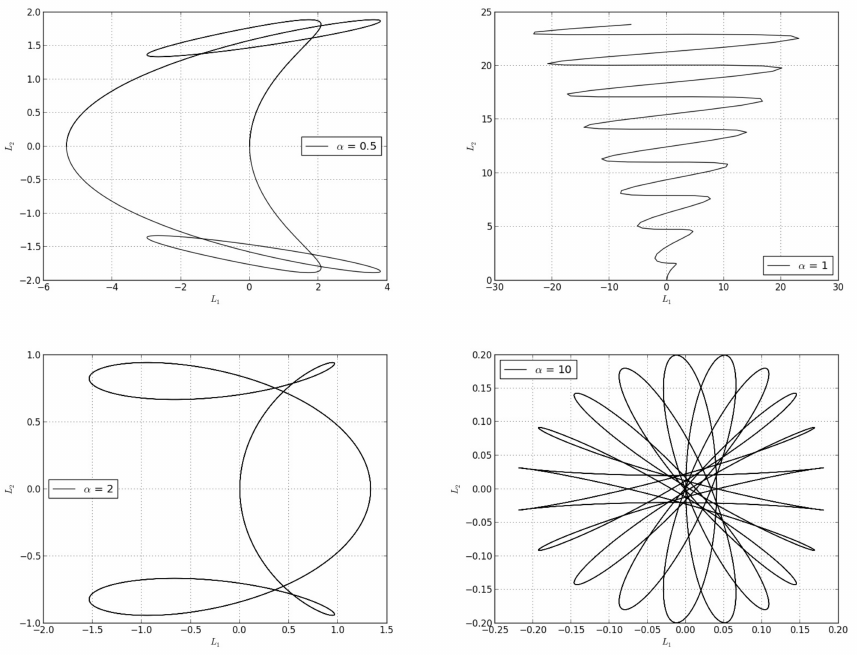}
\end{tabular}
 \label{figur} \caption{Projection of the tip of the angular momentum vector $\mathbf L$ in the $xy$ plane for different values of $\alpha$.}
\end{figure*}

The work-energy theorem states that
\begin{equation}
\int_{\mathbf r_0}^{\mathbf r} \mathbf F\cdot d\mathbf r = \frac{1}{2}m{\mathbf v}^2 - \frac{1}{2}m{\mathbf v_0}^2.
\end{equation}
This may be rewritten as
\begin{equation}
\label{eq:int F.v dt}
\int_0^t P\,dt = \frac{1}{2}m{\mathbf v}^2 - \frac{1}{2}m{\mathbf v_0}^2,
\end{equation}
where the power $P$ is defined as
\begin{equation}
P = \mathbf F\cdot\dot{\mathbf r}.
\end{equation}
That is, the integral of the power $P$ expended by a force $\mathbf F$ acting to move a mass $m$ from time $t=0$ to $t$ is equal to the change in the mass's kinetic energy between these times.

In our model, there are two forces acting on the electron: the Coulomb force $\mathbf F_{c}$ and the perturbing force $\mathbf F_{p}$.  But since the sum of these two forces is $m\ddot{\mathbf r}$, as given in Eq.~(\ref{eq:ddot r is -qE_Coul - qE_pert}), then the left side of Eq.~(\ref{eq:int F.v dt}) becomes
\begin{equation}
\label{eq:P is ddot r dot r dt}
P = m\ddot{\mathbf r}\cdot\dot{\mathbf r}.
\end{equation}
We shall use this equation to compute the power absorbed by the atom.

\subsection{Power: Product Form}

To evaluate the left side of Eq.~(\ref{eq:P is ddot r dot r dt}), we first multiply the expressions for $\ddot{\mathbf r}$ and $\dot{\mathbf r}$ in Eqs.~(\ref{eq:ddot r}) and (\ref{eq:dot r}):
\begin{equation}
\ddot{\mathbf r}\dot{\mathbf r} = (-\mathbf e_1\hat r_0\omega_0^2\hat\psi_0 +\lambda\mathbf e_3\ddot s)(\mathbf e_1\hat r_0 i \mathbf e_3 \omega_0\hat\psi_0 +\lambda\mathbf e_3\dot s)
\end{equation}
Distributing the terms, we get
\begin{align}
\label{eq:ddot r dot r expand}
\ddot{\mathbf r}\dot{\mathbf r} &= -\mathbf e_1\mathbf e_1\hat r_0^*\omega_0^2\hat\psi_0^*\hat r_0 i \mathbf e_3 \omega_0\hat\psi_0 +\mathbf e_3\mathbf e_1\ddot s\hat r_0 i \mathbf e_3 \omega_0\hat\psi_0 \nonumber \\
 & \quad \   - \mathbf e_1\mathbf e_3\hat r_0\omega_0^2\hat\psi_0\dot s + \mathbf e_3\mathbf e_3\ddot s\dot s,
\end{align}
after setting $\lambda = 1$.  Since $i = \mathbf e_1\mathbf e_2\mathbf e_3$ and $ i \mathbf e_3  = \mathbf e_1\mathbf e_2$, then Eq.~(\ref{eq:ddot r dot r expand}) reduces to
\begin{equation}
\ddot{\mathbf r}\dot{\mathbf r} = -i\mathbf e_3 r_0^3\omega_0^3 - i\mathbf e_1\dot s \hat r_0\omega_0\hat\psi_0 + i\mathbf e_2\dot s\hat r_0\omega_0^2\hat\psi_0 + \ddot s\dot s.
\end{equation}
Separating the scalar and bivector parts, we get
\begin{subequations}
\begin{align}
\label{eq:ddot r . dot r}
\ddot{\mathbf r}\cdot\dot{\mathbf r} &= \ddot s\dot s,\\
\ddot{\mathbf r}\times\dot{\mathbf r} &= -\mathbf e_3 r_0^3\omega_0^3 - \mathbf e_1\dot s \hat r_0\omega_0\hat\psi_0 + \mathbf e_2\dot s\hat r_0\omega_0^2\hat\psi_0,
\end{align}
\end{subequations}
after factoring out $i$ in the second equation.

Equation~(\ref{eq:ddot r . dot r}) leads to a very simple expression for the power absorbed by the atom:
\begin{equation}
\label{eq:P is ddot s dot s}
P = m\ddot{\mathbf r}\cdot\dot{\mathbf r} = m\ddot s\dot s.
\end{equation}
Taking the time derivative of $\dot s$ in Eq.~(\ref{eq:dot s final}),
\begin{equation}
\ddot s = \frac{qE_0}{m}\frac{1}{(\alpha^2 - 1)}(-\alpha^2\cos(\alpha\omega_0 t +\varphi) +\cos(\omega_0 t + \varphi)),
\end{equation}
and substituting this and that of $\dot s$ to Eq.~(\ref{eq:P is ddot s dot s}), we get
\begin{align}
P &= \frac{q^2E_0^2}{2m\omega_0^2}\frac{1}{(\alpha^2 - 1)^2}\times \nonumber \\
&\qquad(-\alpha^2\cos(\alpha\omega_0 t +\varphi) + \cos(\omega_0 t + \varphi)) \times \nonumber \\
&\qquad(-\alpha\sin(\alpha\omega_0 t + \varphi)+ \sin(\omega_0 t + \varphi)).
\end{align}
We can show that this is equivalent to 
\begin{align}
\label{eq:P final harmonic}
P &= \frac{q^2E_0^2}{2m\omega_0}\frac{1}{(\alpha^2 - 1)^2}\left(\alpha^3\sin(2\alpha\omega_0t + 2\varphi)\right.\nonumber\\
&\qquad\qquad \left.- (\alpha^2 +\alpha)\sin((\alpha + 1)\omega_0t + 2\varphi)\right.\nonumber\\
&\qquad\qquad +\left. (\alpha^2 -\alpha)\sin((\alpha -1)\omega_0t) \right.\nonumber\\
&\qquad\qquad +\left.+ \sin(2\omega_0t +2\varphi)\right),
\end{align}
which is the desired harmonic form of the power $P$ absorbed by the atom. Notice that power is not constant but fluctuating in time.

\subsection{Average Power over Perturbing Period}

\begin{figure}
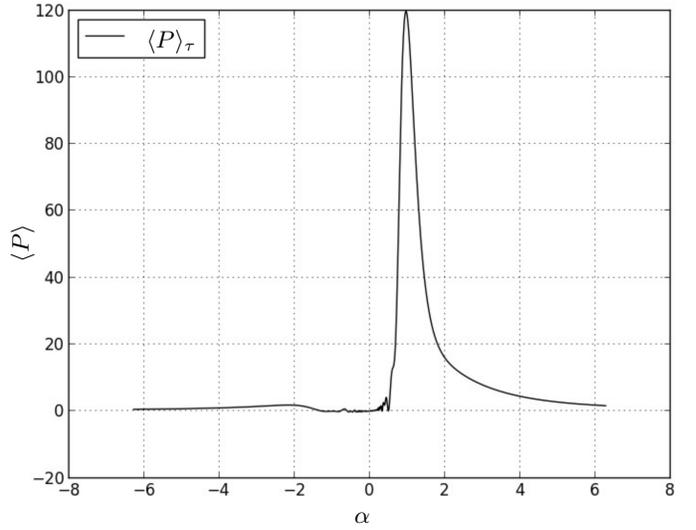

 \centering
   \begin{overpic}[
width=\columnwidth,
tics=5,
page=7
]{rydberg_version2_graphics}
     \put (18,70) {$\langle P \rangle_\tau$}
     \put (50,-3) {$\alpha$}
     \put(-1,40){\makebox(0,0){\rotatebox{90}{$\langle P \rangle$}}}
  \end{overpic}
\caption{This is the graph of the average power over the perturbing period}
 \label{fig:powertau}
\end{figure}

Let us define the average power over the perturbing period as
\begin{equation}
\langle P\rangle_\tau = \frac{1}{\tau}\int_0^{\tau}P\,dt,
\end{equation}
where
\begin{equation}
\tau = \frac{2\pi}{\omega} = \frac{2\pi}{\alpha\omega_0}.
\end{equation}

Now, let us take the time average of the power $P$ in Eq.~(\ref{eq:P final harmonic}) over the perturbing period:
\begin{align}
\label{eq:P final harmonic}
\langle P\rangle_\tau &= \frac{q^2E_0^2}{2m\omega_0}\frac{1}{(\alpha^2 - 1)^2}\left(\alpha^3\langle\sin(2\alpha\omega_0t + 2\varphi)\rangle_\tau \right. \nonumber\\ 
&\qquad \left.- (\alpha^2 +\alpha)\langle\sin((\alpha + 1)\omega_0t + 2\varphi)\rangle_\tau\right.\nonumber\\
&\qquad +\left. (\alpha^2 -\alpha)\langle\sin((\alpha -1)\omega_0t)\rangle_\tau \right.\nonumber \\
&\qquad \left.+ \langle\sin(2\omega_0t +2\varphi)\rangle_\tau\right),
\end{align}
where
\begin{subequations}
\begin{align}
&\langle\sin(2\alpha\omega_0t + 2\varphi)\rangle_\tau \nonumber \\
&\qquad  = 0,\\
&\langle\sin((\alpha + 1)\omega_0t + 2\varphi)\rangle_\tau  \nonumber\\ 
&\qquad = -\frac{1}{2\pi}\frac{\alpha}{(\alpha + 1)}  (\cos((1 + 1/\alpha)2\pi + 2\varphi) - \cos(2\varphi)) \\
&\langle\sin((\alpha -1)\omega_0t)\rangle_\tau  \nonumber \\
&\qquad = -\frac{1}{2\pi}\frac{\alpha}{(\alpha - 1)}(\cos((1 - 1/\alpha)2\pi) - 1),\\
&\langle\sin(2\omega_0t +2\varphi)\rangle_\tau\nonumber \\
&\qquad  = \frac{1}{2\pi}\alpha(\cos(4\pi/\alpha + 2\varphi) - \cos(2\varphi)).
\end{align}
\end{subequations}
Substituting these back to Eq.~(\ref{eq:P final harmonic}), we get
\begin{align}
\langle P\rangle_\tau &= \frac{q^2E_0^2}{4\pi m\omega_0}\frac{1}{(\alpha^2 - 1)^2}\times \nonumber \\
&\quad (-\alpha^2(\cos((1 + 1/\alpha)2\pi + 2\varphi) - \cos(2\varphi))\nonumber\\
&\quad-\alpha^2(\cos((1 - 1/\alpha)2\pi) - 1) \nonumber\\
&\quad +\alpha(\cos(4\pi/\alpha + 2\varphi) - \cos(2\varphi)).
\end{align}

If $\varphi=0$, we have
\begin{align}
\label{eq:avePowerfinal}
\langle P\rangle_\tau &= \frac{q^2E_0^2}{4\pi m\omega_0}\frac{1}{(\alpha^2 - 1)^2}\times \nonumber \\
&\quad (-\alpha^2(\cos((1 + 1/\alpha)2\pi) - 1)\nonumber\\
&\quad -\alpha^2(\cos((1 - 1/\alpha)2\pi) - 1)\nonumber \\
&\quad +\alpha(\cos(4\pi/\alpha) - 1)) .
\end{align}
Equation (\ref{eq:avePowerfinal}) is graphed in Fig. \ref{fig:powertau}. Notice that despite the small oscillations in the interval $\alpha = 0$ and $\alpha = 1$, the power absorption curve is similar to that in Lorentz dispersion theory.

\subsection{Average Power over Orbital Period}

\begin{figure}
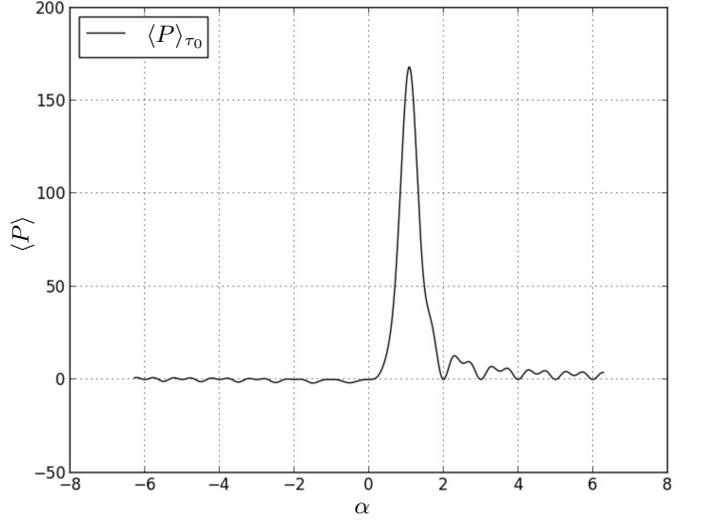

 \centering
   \begin{overpic}[
width=\columnwidth,
tics=5,
page=8
]{rydberg_version2_graphics}
     \put (18,69.5) {$\langle P \rangle_{\tau_0}$}
     \put (50,-3) {$\alpha$}
     \put(-1,40){\makebox(0,0){\rotatebox{90}{$\langle P \rangle$}}}
  \end{overpic}
\caption{This is the graph of the average power over the orbital period}
 \label{fig:powertauo}
\end{figure}

Let us define the average power over the perturbing period as
\begin{equation}
\langle P\rangle_{\tau_0} = \frac{1}{\tau_0}\int_0^{\tau_0}P\,dt,
\end{equation}
where
\begin{equation}
\tau_0 = \frac{2\pi}{\omega_0} .
\end{equation}

Now, let us take the time average of the power $P$ in Eq.~(\ref{eq:P final harmonic}) over the orbital period:
\begin{align}
\label{eq:P final harmonic}
\langle P\rangle_{\tau_0} &= \frac{q^2E_0^2}{2m\omega_0}\frac{1}{(\alpha^2 - 1)^2}(\alpha^3\langle\sin(2\alpha\omega_0t + 2\varphi)\rangle_{\tau_0} \nonumber \\
&- (\alpha^2 +\alpha)\langle\sin((\alpha + 1)\omega_0t + 2\varphi)\rangle_{\tau_0} \nonumber\\
& + (\alpha^2 -\alpha)\langle\sin((\alpha -1)\omega_0t)\rangle_{\tau_0} \nonumber \\
& + \langle\sin(2\omega_0t+2\varphi)\rangle_{\tau_0}),
\end{align}
where
\begin{subequations}
\begin{align}
&\langle\sin(2\alpha\omega_0t + 2\varphi)\rangle_{\tau_0} \nonumber \\
&\qquad= -\frac{1}{2\pi}\frac{\cos(4\pi\alpha+2\varphi)-\cos(2\varphi)}{2\alpha},\\
&\langle\sin((\alpha + 1)\omega_0t + 2\varphi)\rangle_{\tau_0} \nonumber\\
&\qquad= -\frac{1}{2\pi} \frac{\cos(2\pi(\alpha +1))-\cos(2\varphi)}{\alpha+1},\\
&\langle\sin((\alpha -1)\omega_0t)\rangle_{\tau_0} \nonumber \\
&\qquad= -\frac{1}{2\pi}\frac{\cos(2\pi\alpha)-1}{(\alpha-1)},\\
&\langle\sin(2\omega_0t +2\varphi)\rangle_{\tau_0} \nonumber \\
&\qquad= 0.
\end{align}
\end{subequations}
Substituting these back to Eq.~(\ref{eq:P final harmonic}), we get
\begin{align}
\langle P\rangle_{\tau_0} &= \frac{q^2E_0^2}{8\pi m\omega_0}\frac{1}{(\alpha^2 - 1)^2}\times \nonumber \\
& \qquad \quad (-\alpha^2(\cos(4\pi\alpha + 2\varphi)-\cos(2\varphi))  \nonumber\\
&\qquad\quad -2\alpha(\cos(2\pi(\alpha + 1)) - \cos(2\varphi)) \nonumber \\
&\qquad \quad - 2\alpha (\cos(2\pi\alpha)-1)) .
\end{align}

If $\varphi=0$, we have
\begin{align}
\label{eq:avePoweroverOrbitalPeriod}
\langle P\rangle_{\tau_0} &= \frac{q^2E_0^2}{8\pi m\omega_0}\frac{-1}{(\alpha^2 - 1)^2}(\alpha^2(\cos(4\pi\alpha)-1)  \nonumber\\
& -2\alpha(\cos(2\pi(\alpha + 1)) - 1) - 2\alpha (\cos(2\pi\alpha)-1)).
\end{align}
Equation (\ref{eq:avePoweroverOrbitalPeriod}) is graphed in Fig. \ref{fig:powertauo}. Notice that unlike in Fig. \ref{fig:powertau}, there are periodic oscillations after $\alpha=2$. 

\section{Conclusion and Recommendation}

We modelled the classical Hydrogen atom as an electron revolving in circular orbit around an immovable proton subject to the Coulomb force. We subjected this atom to a perturbing oscillating electric field perpendicular to the electron's initial orbital plane. We showed that the resulting equations of motion of the electron along the axis of the perturbing electric field is similar to that of a simple harmonic oscillator with sinusoidal forcing.  Furthermore, the absorbed energy averaged over the period of the perturbing field or over the orbital frequency of the electron is approximately similar to a resonance curve with one dominant frequency with finite peak at $\omega = \omega_o$; other small resonance peaks occur to the left or to the right of the major resonant frequency.  

The Lorentz dispersion model of the light-atom interaction assumes that the electron is subject to Hooke's force and the force due to the oscillating electric field of the light. Interestingly, even if our initial assumption is an electron in circular orbit around the nucleus, we still obtained the same forced harmonic oscillator equation as that of the standard model. We also obtained the same resonant frequency, though the actual peak is at a frequency slightly smaller than $\omega_0$. But what is new is that even though we did not put a damping term in our harmonic oscillator equation, we still obtained a finite energy absorption at the resonant frequency $\omega$. We also computed the electron's angular momentum vector and showed that its tip traces rosette patterns similar to epicycles\cite{gallavotti2001quasi}. 

In the future work, we shall extend our work to the interaction of the hydrogen atom with elliptically polarized radiation.

\section*{Acknowledgements}
This work was supported by the Loyola Schools Scholarly Work Faculty Grants of the Ateneo de Manila University.

\bibliography{rydberg_version2Bib}

\end{document}